# Uncovering sustainable personal care ingredient combinations using scientific modelling

**Bhattacharya, Sandip**[1*]; Freitas, Vanessa da Silva[2]; Kohlmann, Christina[1]

[1] BASF Personal Care and Nutrition GmbH, Düsseldorf, Germany;
[2] BASF South America, Sao Paulo, Brazil

* Corresponding author: Dr. Sandip Bhattacharya, BASF Personal Care and Nutrition GmbH, Z22, Henkelstrasse 67, 40589 Duesseldorf, Germany, +49 160 96713479, and sandip.bhattacharya@basf.com

**Abstract**

Personal care formulations often contain synthetic and non-biodegradable ingredients, such as silicone and mineral oils, which can offer a unique performance. However, due to regulations like the EU ban of Octamethylcyclotetrasiloxane (D4), Decamethyl-cyclopentasiloxane (D5), Dodecamethylcyclohexasiloxane (D6) already in effect for rinse off and for leave on cosmetics by June 2027 coupled with growing consumer awareness and expectations on sustainability, personal care brands face significant pressure to replace these synthetic ingredients with natural alternatives without compromising performance and cost. As a result, formulators are confronted with the challenge to find natural-based solutions within a short timeframe. In this study, we propose a pioneering approach that utilizes predicting modelling and simulation-based digital services to obtain natural-based ingredient combinations as recommendations to commonly used synthetic ingredients. We will demonstrate the effectiveness of our predictions through the application of these proposals in specific formulations. By offering a platform of digital services, it is aimed to empower formulators to explore good performing novel and environmentally friendly alternatives, ultimately driving a substantial and genuine transformation in the personal care industry.

**Introduction**.

The personal care industry belongs to the Fast-Moving Consumer Goods (FMCG) market, with over 150,000 new consumer product launches per year [1,2]. Significant number of consumers are increasingly making their purchase decisions based on performance & claims of the underlying ingredients by doing their research beforehand. The top 10 brands represent a little below 20% of the market, meaning there is a large share of mid-sized and indie brands in the personal care sector, too. Furthermore, several product sub-categories are driving complexity. Thus, there is a constant need to innovate for players who want to be active in this consumer centric business. The statistics mentioned here have been inspired from an extensive review on the state of the beauty market made by Mckinsey [2].

Currently, social-media has attributed to a wide-spread consumer awareness in terms of sustainability. This means that customers want their cosmetics to be based on natural ingredients, but with the same performance as standard cosmetics which are usually comprised of well-established performing, but synthetic ingredients. At the same time in the interest of the business, the additional cost of replacing synthetic with natural ingredients should be minimal.

Hence, the problem of replacing synthetic ingredients is not trivial. Let's say, we are talking about exchanging silicones or mineral oils in skin-care formulations. The solution to uncover the best natural alternatives is like finding a needle in a haystack. The conventional formulator's approach to identifying replacements to a particular cosmetic ingredient is to rely on their expertise for solutions and to test it in the formulation on a case-by-case, trial-error approach. In most cases, this is a time consuming and costly approach, with the risk of missing out possible good performing (untested) solutions. It would be extremely valuable to a personal care formulator and profitable for personal care companies to obtain quick solutions to the needle in a haystack problem, so that they can focus only on testing the best performing natural-based solutions in the formulation.

Modelling and simulation based on digital services could provide such solutions to the industry. Furthermore, digital services are boosted by constantly increasing computing powers (supercomputers and prospect of wide-spread use of quantum computers in the future) and accurate easy to use prediction methods (scientific and mathematical modelling and machine learning). Similar methodologies are already established to identify new molecules with desired performance in the pharmaceuticals industry [3] among many others.

In this paper, we discuss the details of scientific modelling based digital services that provide natural based solutions to tackle key sustainability challenges faced by the industry. Like a human brain using short-cuts (heuristics) to make key decisions required for our survival [4,5], in this work we will demonstrate our modelling and artificial intelligence skill sets relying on careful approximations to obtain solutions, with the ultimate goal of supporting a sustainable development of the personal care industry.

**Methods**.

*Eq. 1*

$$prop = prop_{signal} + prop_{noise}$$

An essential cornerstone to statistics is that if the measurement of a property (*prop*) is repeated, it may give slightly different values. Often in our reporting, we have a metric of this error, for example as $R^2$ error or standard deviation among others. All data can be described as a linear summation of a signal and a noise part. While the signal delivers essential intrinsic information to the property (*prop*) that is measured and is theoretically fixed when the experiment is repeated, the noise part can be random upon repetition and is dictated by several features like experimental error, environmental influences (temperature or humidity) or personal influences, lack of standardization, if the experiment is repeated after a pause, etc. Eq. 1 is valid for all properties either measured by humans e.g., sensory assessment or by machines, like a conventional rheometer measuring the viscosity of a fluid. Trained sensory panelists would strive to reduce the noise component of a property which is measured (sensory attributes) through the training procedure. Likewise, a working rheometer would have a proportionally lower amount of the random noise in its output data, which can also be termed as the experimental error. In our work we always strive to uncover the simplest model that delivers a good score on previously unseen data i.e., data which the model has not used in training. The simplicity in the prediction models is key to prevent overfitting, i.e., the model mistakeably fitting the noise instead of the signal component of the data (Eq. 1). Observations and properties which have a dominant noise-component of the data cannot be modelled by the methods illustrated in this paper. However, a wide variety of pain-points of the FMCG business can be addressed through modelling and simulation based digital services covered here.

We will describe the different modelling methods used in this work to recommend alternatives to synthetic ingredients in cosmetics formulations. To make this section understandable to those beyond just mathematicians or statisticians, we have decided to include only the necessary amount of mathematical equations in this section.

**Simple mathematical modelling**

Let us assume an ingredient i, used in a personal care formulation. It can be an ingredient present in high concentrations like an emollient or a surfactant, dictating the properties of a leave on or a rinse of formulation, respectively. Indeed, even within a trial-and-error approach, the starting point of many personal care formulators, is based on the properties of the individual ingredients ($prop^i$), when deciding which combinations of ingredients (e.g., i= specific emollients or surfactants) to test. For emollients $prop^i$ can be properties measured by physico-chemical methods e.g., like polarity (the water-oil interfacial tension, IFT), viscosity, the air-oil surface tension (SFT), refractive index (RI), alkane carbon number (ACN) or spreading value. Likewise, this can also be monadically assessed sensory attributes by panelists e.g., powdery, waxy, dry, silicone-oil feel, etc. Similarly, for surfactants, $prop^i$ can be properties measured by physico-chemical methods e.g., like the (dynamic) time dependent foam heights, interfacial tension, surface tension, foam elastic structure, bubble size or again sensory properties evaluated by trained panelists.

Personal care formulations contain mixtures of different ingredients. For example, a formulator can combine a light, medium and a heavy emollient to mimic a specific sensory profile, say of a reference containing silicones. However, when starting from scratch, which emollient (i) to take in which ratio ($x^i$), is the concrete needle in a haystack problem. The mathematical models used in the predictions to obtain the properties of the mixture, $prop^{mix}$, are of the following nature:

*Eq. 2*

$$prop^{mix} = f(x^n, prop^n)$$

Here, mixture (mix) can be that of an emollient mixture, i.e. those between emollient labelled as i, j, k... In the above equation *f* is the desired mathematical prediction model that is a summation over the components of the mixtures (i,j,k). The function *f*, i.e. the prediction model, can be a polynomial, exponential or a logarithmic one [6]. In the present work, the accuracy of these mixing models was more than 95% on (untrained) test experimental data [6].

**Target or benchmark optimization**

Next, the cosmetic formulator needs to use the different prediction mixing models in Eq. 1 for the numerous properties discussed above to obtain recommendations for personal care ingredients that allow the best match to a given benchmark (like silicone or mineral oils). This translates to an optimization problem in mathematics, i.e. to find the ratios $x^n$ and identify the components (i,j,k…..of the mixture above) that minimizes the following error:

*Eq. 3*

$$\Delta = |prop^{benchmark} - prop^{mix}|^2$$

Such an optimization is like solving a least squares problem [8,9] to obtain $x^n$ (ratio) and n (the components i, j, k) that returns the smallest possible error in the above equation Eq. 1. Note that in Eq. 2, we have a vector of properties $\boldsymbol{prop^{benchmark}}, \boldsymbol{prop^{mix}}$, which include the physico-chemical and sensory properties discussed above. Additionally, our optimization models go through all possible ingredient (e.g. emollient or surfactant) combinations and return the best matching ingredient combinations to a benchmark (e.g. silicone oil or synthetic sulfated surfactant). Therefore, a formulator, will receive best theoretical solutions to match a benchmark, i.e., they have successfully identified the needle in the haystack! In this work, we have come up with two simulators namely Emollient and Surfactant Simulators to uncover natural-based replacements to synthetic oils and surfactants respectively, applicable in skin, hair and oral care applications.

**Machine learning**

*Eq. 4*

$$prop^{formulation} = f_{ML}(\boldsymbol{x^n}, \boldsymbol{prop^{mix}})$$

For more complex problems such as formulation modelling, where several different types of ingredients are involved and simple mathematical modelling is not possible, we use machine learning models. Here $prop^{formulation}$, can be stability of a formulation, its physico-chemical or sensory properties. The above equation has vectors $(\boldsymbol{x^n}, \boldsymbol{prop^{mix}})$ which simply indicate the ratios and properties respectively running over all ingredients in the formulation.

The machine learning models $f_{ML}$, have been constructed in Python using standard available packages [7]. The approach involves testing a basket of models on the available data and choosing the models returning the best score on the test (unseen) dataset. For more complex chemistry-property predictions, we have also used advanced machine learning methods like Graph Neural Networks [10,11].

It is important to highlight that with the above approaches we have made several scientific approximations, for example ingredient descriptors indicating the complex mechanisms and interactions in a final formulation. The inspiration for this is the functioning of a human brain. Note that a literal solution to several complex decision-making problems that we encounter daily (i.e., the most accurate solution), would mean consumption of a large amount of energy and time. As a direct evolutionary consequence our brains make short-cuts which provide an approximate and an almost correct solution to multiple problems [4,5]. Undoubtedly, our research endeavors to develop similar computational intelligence-based digital approaches aiming to address the challenges faced by the personal care industry.

Next, we demonstrate the prediction abilities of our models to solve daily-life challenges of the beauty industry.

**Results and Discussion**.

In this section we will provide examples i.e., solutions obtained from the modelling methodologies described previously, which can be directly incorporated to generate sustainable formulation ingredients.

**Natural replacements to Cyclopentasiloxane**

Certain cosmetics formulations for skin care, hair oils and color cosmetics (like makeup removals) use large quantities of cyclosiloxanes emollients, D4, D5 and D6. Out of these predominantly D5 is used extensively due to its good stability, volatility and sensory properties in combination with its low cost. D4, D5 and D6 were identified by the Member State Committee (MSC) of the European Chemicals Agency ('the Agency') as Substances of Very High Concern (SVHC) with very Persistent and very Bioaccumulative (vPvB) properties. D4 was identified as having Persistent, Bioaccumulative and Toxic (PBT) properties. D5 and D6 were also identified as having PBT properties when they contain 0.1 % or more by weight of D4. More details can be found in European Commission's regulations to amend the REACH Regulation (Annex XVII) concerning these silicones [12].

Additionally, now with official regulations already published regarding authorization and restriction of this substance, personal care companies must replace the mentioned silicones with other alternatives. They are urgently seeking a complete replacement for these cyclic silicones in particular D5, with natural alternatives for every application.

*Figure 1*

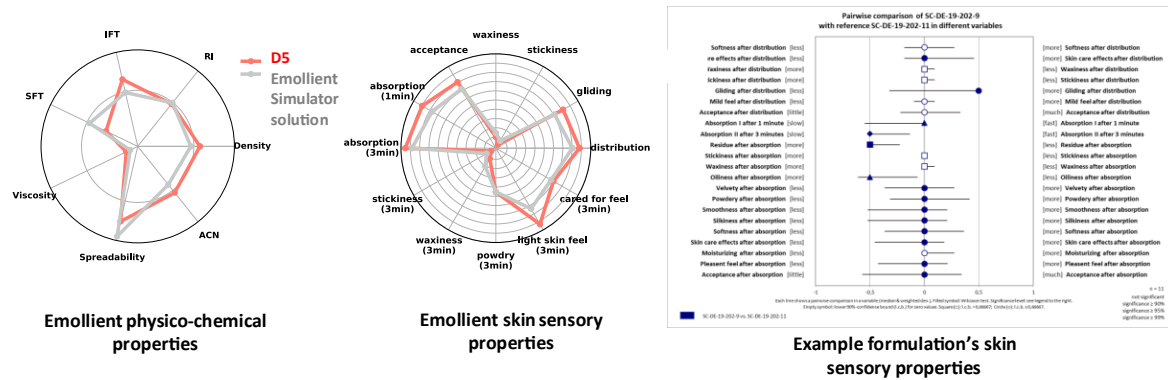

Emollient physico-chemical properties | Emollient skin sensory properties | Example formulation's skin sensory properties

In Figure 1, we demonstrate how Emollient Simulator can be used to uncover natural-based replacements to D5 for a specific make-up cleansing application. Based on the methods described above we obtain a proposed *modelled* ternary mixture of coco-caprylate/caprate, undecane & tridecane and dicaprylyl ether in an optimized weight ratio 1:2:1 of the three commercial emollients. Comparison between the physico-chemical and monadic sensory properties of the D5 (measured) vs. the Emollient Simulator solution (predicted) is illustrated in the left and middle panels of Figure 1, illustrating a good agreement at the emollient level between the synthetic D5 and our solution derived from 100% natural feedstocks [15].

On the right panel there is a sensory-pair wise comparison of a simplex makeup removal formulation with D5 (at baseline 0) and the one containing the proposed mix from Emollient Simulator. In addition to the emollient, the simplex formulation contains water, NaCl, complexing agent and alcohols. The pairwise comparisons, exhibit a good agreement between the two formulations as indicated by the majority of the sensory attributes lying at the baseline. Additionally, we have investigated that the simplex formulation with natural-based solutions delivers similar cleansing and makeup (mascara, lipstick and foundations) removal performance. This illustrates that the Emollient Simulator's predictions could be used as a drop-in replacement solution for different formulations.

**Natural replacements to Dimethicones**

Personal care cosmetics also widely use other silicones like linear polydimethylsiloxane or dimethicones. Although, dimethicones are not banned, similar regulations to cyclosiloxanes are anticipated in the future. Moreover, consumers want more silicone-free formulations, translating to a rapid launch of such market products and simultaneously a decrease in dimethicones in skin care, in the recent years [1]. Polydimethylsiloxane is essentially like a polymer with the degree of polymerization dictating the product viscosity and therefore application. Lighter dimethicones (1 to approximately 250 cSt) have emollient-like properties and are used in skin care applications, while heavier ones (> 1000 cSt) have gum or polymer like properties and are used in hair care applications. Although these dimethicones have different structure and properties depending on their viscosities, they appear in the INCI of a market formulation as *dimethicone.* This results in an additional layer of complexity in replacing this family of silicones in cosmetic formulations.

In skin care, lighter dimethicones (< 50 cSt) are used for their sensorial properties in cosmetics applications, while heavier ones (> 50 cSt) are used primarily for their anti-foaming properties in addition to their sensorial properties. In the interest of controlling the length of this paper, we will show just one example, how Emollient Simulator can be used to obtain natural-based replacement to a specific dimethicone of viscosity 5 cSt. Our digital methodology can be used to provide accurate natural-based solutions to dimethicones with viscosities 1-200 cSt.

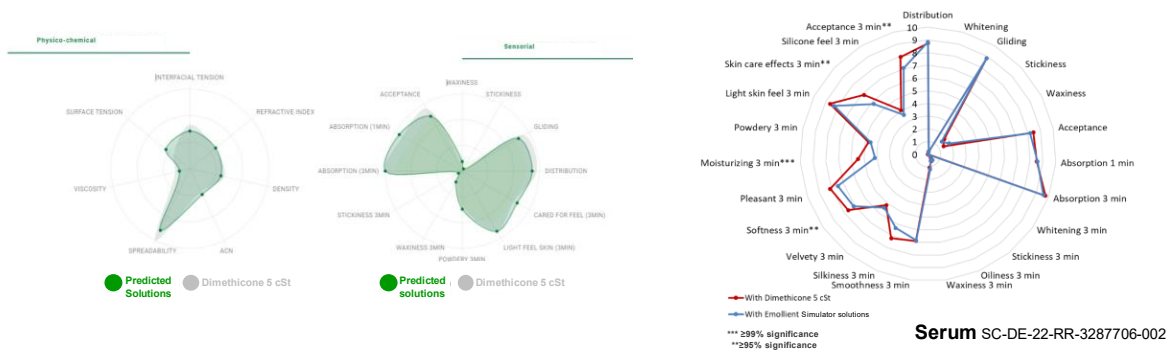

The left and middle panel of Figure 2 shows the physico-chemical and monadic sensory comparisons between the experimentally measured Dimethicone 5 cSt and a natural-based ternary mixture of dicaprylyl carbonate, dicaprylyl ether and undecane/tridecane (1.7:3.2:1) predicted from Emollient Simulator, showing a good agreement between the silicone and natural-based mixtures. The right panel above shows monadic pairwise comparison between a reference skin-care serum formulation containing dimethicone 5 cSt versus the natural-based emollient mixture obtained from above, exhibiting a good agreement also at the formulation level. The simplex formulation contains complexing agents, preservative, O/W emulsifier, rheology modifier, citric acid and water in addition to a significant emollient part.

Aside from dimethicones being a sensory emollient, particularly the heavier ones (> 50 cSt) perform a dual role in acting also as an anti-foaming agent. Other components in common natural skin care formulations like the fatty alcohol-based consistency agent or xanthan gum have been known to produce foaming when rubbed intensively on skin by a user. This creates an undesired sensory effect in dimethicone-free formulation commonly found in the present market. To remove this effect, in addition to replacing heavier dimethicones by natural-based solutions obtained from Emollient Simulator, one needs to adjust other components of the formulations. The simplex serum formulation used above does not exhibit any foaming effect also with the natural-based emollient mixture, i.e. without dimethicone.

**Natural surfactant replacement to synthetic α-olefin sulfonate**

Finally, we demonstrate the use of Surfactant Simulator with an example to obtain natural-based solutions to a synthetic surfactant. α-olefin sulfonate is a commonly used anionic surfactant in rinse-off applications like shampoos, hand and body wash formulations. However, it is comprised of synthetic feedstocks. Often α-olefin sulfonate is used in combination with a small amount of amphoteric cocamidopropyl betaine, which is partly natural-based. The combination of α-olefin sulfonate and cocamidopropyl betaine presents a good foaming performance. However, the large quantities of synthetic α-olefin sulfonate make it unsuitable to the demand of natural cosmetics.

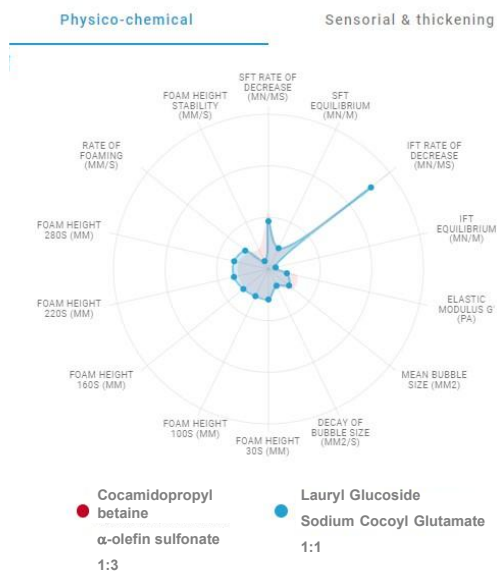

From the computational methodologies described before the Surfactant Simulator can be used to uncover natural, non-ethylene oxide and non-sulfated surfactant-based alternatives to a mixture of α-olefin sulfonate and cocamidopropyl betaine (with an active surfactant matter weight ratio of 3:1), the commonly used mixture in rinse-off formulation with the anionic surfactant. The best performing solution is a mixture containing a natural-based lauryl glucoside (an alkylpolyglucoside surfactant) and sodium cocoyl glutamate (an amino-acid surfactant), with an active surfactant matter weight ratio of 1:1. Indeed the performance properties like foaming, bubble formation and elastic coefficient of foam of the natural based mixture of lauryl glucoside and sodium cocoyl glutamate is very close to that of the mixture containing synthetic α-olefin sulfonate.

Again, with the interest of controlling the length of this paper, we have demonstrated only one example use from Surfactant Simulator. The digital models behind Surfactant Simulator can be used to obtain natural-based solutions to a wide variety of synthetic surfactant mixtures.

**Conclusion**

In this study, we have delved into the potential of modelling and simulation-based digital services to empower formulators who are tasked with discovering sustainable alternatives to synthetic ingredients in personal care products. With the increasing demand for sustainable alternatives to substances like silicones, mineral oils, and sulfated surfactants, formulators are faced with a difficult, expensive, and time-consuming challenge.

It was extensively discussed how prediction models at the ingredient level and machine learning models can serve as descriptors to provide insights into the performance of ingredients in formulations. Specifically, we have focused on three examples: cyclopentasiloxane, a specific dimethicone and α-olefin sulfonate, aiming to offer natural-based alternatives for these ingredients. Our findings demonstrate that the formulation performance of our ingredient level recommended natural-based solutions are comparable to formulations containing the respective synthetic ingredients.

Moreover, the computational methodology outlined in this research can be further applied to identify sustainable solutions for various other synthetic ingredients commonly used in personal care products, such as isododecane, isohexadecane, mineral oils, squalene, among emollients, and taurates, sodium laureth sulfate, isethionates, sarcosinates, among surfactants. By utilizing these digital services, the personal care industry can significantly reduce formulation development times, offering substantial benefits.

**Acknowledgments.**
This work was sponsored by BASF.

**Conflict of Interest Statement**.
NONE